\documentclass{appolb}
\usepackage{graphicx}
\usepackage{amsmath}

\usepackage{lineno}
\begin{document}
\title{From Elastic Scattering to Central Exclusive Production: Physics with Forward Protons at RHIC%
\thanks{Presented at 60th Cracow School of Theoretical Physics, Acta Phys. Pol. B 52, 217 (2021)}
\thanks{This work was supported by the U.S. Department of Energy under contract number de-sc0012704}%
}
\author{W\l odek Guryn
\address{Brookhaven National Laboratory}
}
\maketitle
\begin{abstract}
We describe a physics program at the Relativistic Heavy Ion Collider (RHIC) with tagged forward protons.
The program started with the proton-proton elastic scattering experiment (PP2PP), for which a set of Roman Pot 
stations was build. The PP2PP experiment took data at RHIC as a dedicated experiment at the beginning of RHIC operations.
To expand the physics program to include non-elastic channels with forward protons, like Central Exclusive Production (CEP), Central Production (CP) and Single Diffraction Dissociation (SD), 
the experiment with its equipment was merged with the STAR experiment at RHIC.
Consequently the expanded program, which included both elastic and inelastic channels became part of the physics program and operations of  the STAR experiment.
In this paper we shall describe the physics results obtained by the PP2PP and STAR experiments to date.
\end{abstract}
  
\section{Introduction}
In the description of the physics program we shall follow the time line and evolution of the program from its inception as a PP2PP experiment~\cite{bib:BultmannNIM2004} to the most recent results with the STAR experiment~\cite{bib:STARNIM}. 
We start by introducing basic quantities needed to describe elastic scattering.

\begin{figure}[htb]
\centerline{
\includegraphics[width=0.3\textwidth]{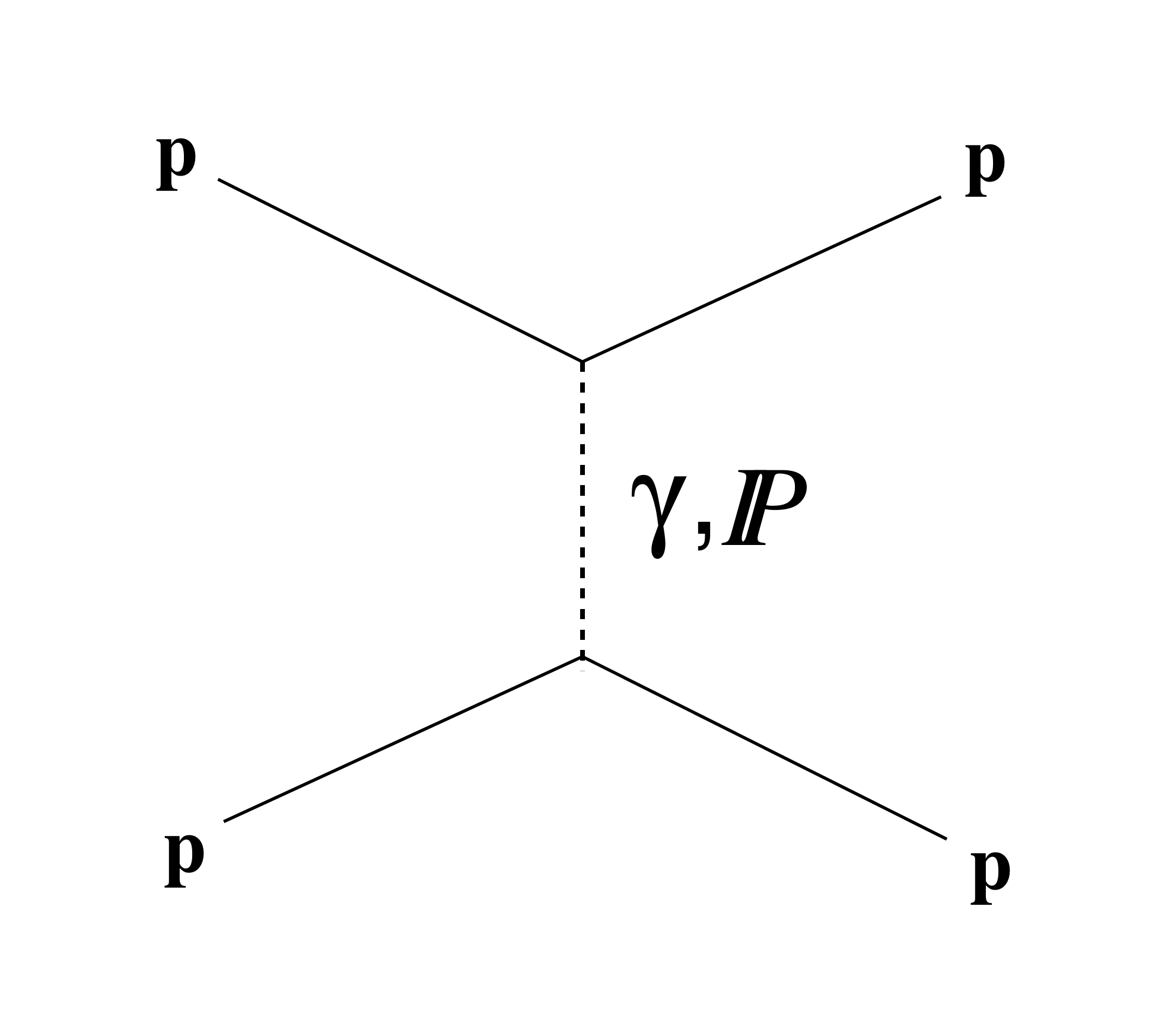}}
\caption{Diagram of $pp$ elastic scattering.}
\label{fig:ETdiagram2}
\end{figure}

The proton-proton elastic scattering, Fig. \ref{fig:ETdiagram2} is described by a scattering amplitude which has two components:
the electromagnetic part, described by the well-known Coulomb amplitude $f_c$, and the hadronic part, described by the hadronic amplitude $f_h$. 
The hadronic amplitude is commonly expressed as function of center-of-mass energy $\sqrt s$ and four-momentum-transfer squared $t$, which for small scattering angles $\theta$ can be expressed as:

\begin{equation}
t = (p_{in} - p_{out} )^2 \approx - p^2\theta^2 = - p^2\cdot ( \theta_x^2 + \theta_y^2).
\label{eq:t-def}
\end{equation}

The differential elastic $pp$ cross section is expressed as a square of the scattering amplitude:

\begin{equation}
\frac{d\sigma_{el}}{dt}  = \pi \vert f_c + f_h \vert ^2.  \label{eq:dsdt}
\end{equation}

Thus, the differential cross section $d\sigma_{el}$/d$t$ for elastic scattering in the forward angle region is determined 
by Coulomb and nuclear amplitudes and the interference term between them. The cross section is given by~\cite{bib:amaldi}:

\begin{equation}
\frac{{\rm d} \sigma_{el}}{{\rm d} t}  = 4\, \pi\, (\hbar c)^2 \left( \frac{\alpha \, G_E^2}{t} \right)^2 + \frac{1 + \rho^2}{16 \, \pi \, (\hbar c)^2} \cdot \sigma_{tot}^2 \cdot e^{-B \,|t|} \nonumber - (\rho + \Delta \Phi) \cdot \frac{\alpha \, G_E^2}{|t|}  \cdot \sigma_{tot} \cdot  e^{-\frac{1}{2}B \,|t|}
\label{eq:dsigmadt}
\end{equation}
where $\alpha$ the fine structure constant, $G_E$ the electric form factor of the proton, $\Delta \Phi$ the Coulomb phase~\cite{bib:PhaseKopeliovich}, $\rho$ the ratio of the real to imaginary part of the forward scattering amplitude, 
$\sigma_{tot}$ the total cross section, and $B$ the nuclear slope parameter. The first term in Eq.~\ref{eq:dsigmadt} is Coulomb term, the second is the hadronic term and the third is so called Coulomb Nuclear Interference (CNI) term.

At the time of the proposal of the PP2PP experiment the UA4 experiment~\cite{bib:UA41987rho} published an anomalously large $\rho$-value.  Hence it was quite natural to propose an experiment at RHIC  at $\sqrt s = 500$ GeV, which among other things, could measure the $\rho$-value in proton-proton collisions at a similar to UA4 experiment's $\sqrt s = 546$ GeV. Hence a comprehensive experiment to measure total and elastic cross sections was proposed for the RHIC physics program.

\section{Results from PP2PP experiment}
The PP2PP experiment obtained few results on $pp$ elastic scattering at $\sqrt s = 200$~GeV. These were the first results from $pp$ elastic scattering above the ISR $\sqrt s$ of 63 GeV, the highest $\sqrt s$ at the time. The first result was obtained with limited statistics due to the short data taking time, the $B$-slope~\cite{bib:BueltmanPLB2004} parameter was measured in the small $t$-range $0.010 \le -t \le 0.019\: \text{GeV}^2$ to be $B = 16.3 \pm 1.6 (stat.) \pm 0.9 (syst.)\: \text{GeV}^{-2}$. The error bars were rather large because of the limited statistics of the measurement.
Spin dependence in $pp$ elastic scattering was also measured. Both the single spin asymmetry $A_N$~\cite{bib:BueltmanPLB2006} 
and double spin asymmetry $A_{NN}$~\cite{bib:BueltmanPLB2007} were measured in the low-$t$ region, $ 0.002 < -t < 0.03\: \text{GeV}^2$. 
The $A_N$ in this $t$-range is sensitive to a possible contribution from the hadronic spin-flip amplitude, which is due to the 
interference of the Pomeron spin-flip amplitude and electromagnetic non-flip amplitude and would change $A_N$.
The common measure of this effect is the variable $r_5$~\cite{bib:Kopeliovich1989,bib:Buttimore1999}, which measures the ratio of hadronic spin-flip to non-flip amplitudes.
It was found that the $A_N$ follows the Coulomb Nuclear Interference curve and that the $r_5$ fitted to the data was compatible with no hadronic spin-flip. 
The double-spin asymmetry $A_{NN}$ was also found compatible with zero.

\section{Results on elastic scattering at STAR}

As mentioned earlier, the Roman Pot setup of the PP2PP experiment was subsequently moved to and became part of the STAR to continue $pp$ elastic scattering program and also to expand physics program to include diffractive scattering with tagged forward protons. We start with the results on elastic $pp$ scattering published to date, where two major results were obtained. 

\begin{figure}[htb]
\centerline{
\includegraphics[width=0.55\textwidth]{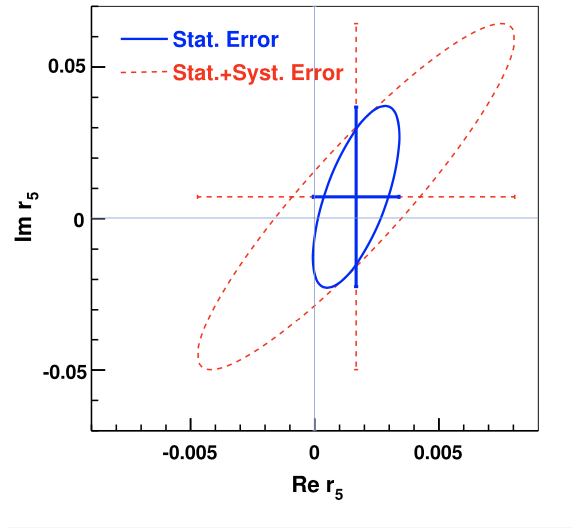}}
\caption{Fitted value of $r_5$, $Im\: r_5$ vs $Re\: r_5$, with contours corresponding to statistical error (solid ellipse and cross) and statistical + systematic errors added in quadrature (dashed ellipse and cross) of $1\sigma$.}
\label{fig:r5200GeV}
\end{figure}

The first was an improved result on the $A_N$, which put a more stringent limit on the hadronic spin-flip amplitude $r_5$ at $\sqrt s = 200$~GeV was obtained~\cite{bib:STARAN2013} using higher statistics
in $t$-range $0.003 \le -t \le 0.035 \: \text{GeV}^2$, 
where there is a significant interference between the electromagnetic and hadronic scattering amplitudes. 
The measured values of $A_N$ and its $t$-dependence are consistent with a vanishing hadronic spin-flip amplitude within $1\sigma$, see Fig.~\ref{fig:r5200GeV}, where the the real and imaginary parts of $r_5$ are plotted.
The measurement provides a strong constraint on the ratio of the single spin-flip to the non-flip amplitudes. Since the hadronic amplitude is dominated by the Pomeron amplitude at this $\sqrt s$, 
we conclude that a strong constraint on the presence of a hadronic spin flip due to the Pomeron exchange in polarized $pp$ elastic scattering was obtained.

\begin{figure}[htb]
\centerline{
\includegraphics[width=0.45\textwidth]{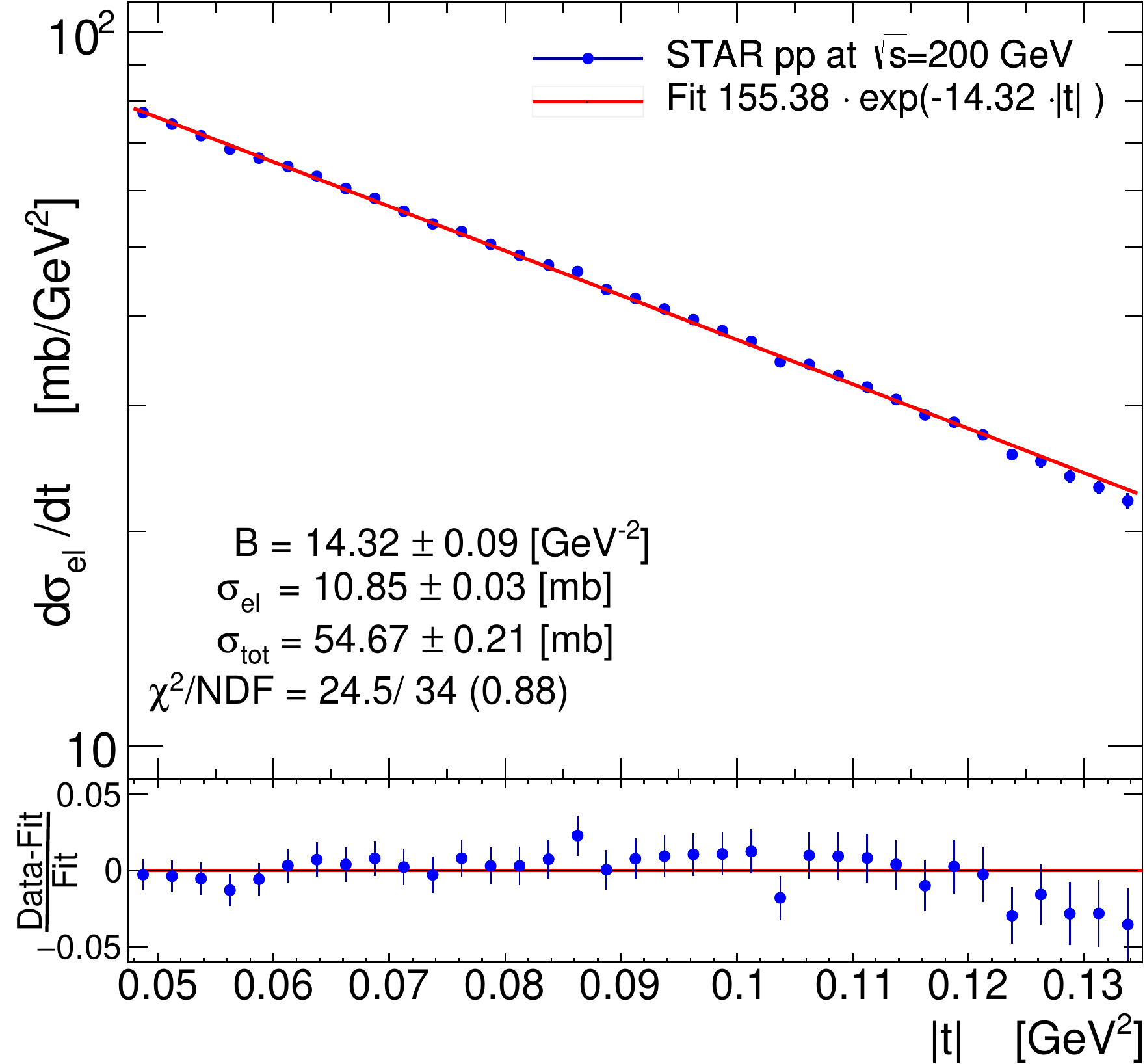}}
\caption{Top panel: $pp$ elastic differential cross-section $d\sigma/dt$ fitted with exponential $A\exp{(Bt)}$, results are shown in the panel; Bottom panel: Residuals (Data - Fit)/Fit. Uncertainties are statistical only.}
\label{fig:Elastic200GeV}
\end{figure}

The second was a measurement of the elastic differential cross section measured in the $t$-range $0.045 \leq -t \leq 0.135$~GeV$^2$~\cite{bib:STARET2020}.
The results are shown in Fig.~\ref{fig:Elastic200GeV} and they included the following observables.
The value of the exponential slope parameter $B$ of the elastic differential cross section $d\sigma/dt \sim e^{-Bt}$ in the measured $t$-range
was found to be $B = 14.32 \pm 0.09 (stat.)^{\scriptstyle +0.13}_{\scriptstyle -0.28} (syst.)$~GeV$^{-2}$. 
The total cross section $\sigma_{tot}$, obtained from extrapolation of the $d\sigma/dt$ to the optical point at $t = 0$, is $\sigma_{tot} = 54.67 \pm 0.21 (stat.) ^{\scriptstyle +1.28}_{\scriptstyle -1.38} (syst.)$~mb.
Obtained were also the values of the elastic cross section $\sigma_{el} = 10.85 \pm 0.03 (stat.) ^{\scriptstyle +0.49}_{\scriptstyle -0.41}(syst.)$~mb,
the elastic cross section integrated within the STAR $t$-range  $\sigma^{det}_{el} = 4.05 \pm 0.01 (stat.) ^{\scriptstyle+0.18}_{\scriptstyle -0.17}(syst.)$ mb, and the
inelastic cross section $\sigma_{inel} = 43.82 \pm 0.21  (stat.) ^{\scriptstyle +1.37}_{\scriptstyle -1.44} (syst.)$ mb. The result on $\sigma_{tot}$ is the only one on the total cross section at $\sqrt s$ between the ISR and the LHC energies. All the results agree well with the world data.

\section{Results on Central Exclusive Production at STAR}
The major motivation to move the Roman Pot system to STAR was a possibility of combining the measurement of the forward protons with the the measurement of the central (recoil) system in the STAR detector, 
thus allowing the determination of the exclusivity of the final state in $pp \rightarrow pXp$ reaction, 
where all the particles in the final are measured. 
The verification of the exclusivity of that final state is a very unique feature of the results described here, as it is very common to infer exclusivity by requiring a rapidity gaps and of a small $p_T$ of the final state. 
The latter is a reasonable approximation of the exclusivity but it is clearly not the same as the measurement of all particles in the final state.
In the results described here, the system $X$ consists of hadron pairs of opposite charge $\pi^+\pi^-$, $K^+K^-$ and $p \bar p$.
We present a brief summary of the results obtained at $\sqrt s = 200$~GeV, for a more detailed discussion see~\cite{bib:STARCEP2020}.

The charged particle pairs' momenta were obtained from the tracks measured in the Time Projection Chamber (TPC) of the central detector. The particle identification (PID) was obtained using the energy loss ($dE/dx$) measured in the TPC and the time of flight measurement in the Time of Flight (TOF) system.
The forward-scattered protons were reconstructed in the Roman Pot system. Exclusivity of the event  was determined by requiring the transverse momentum balance of all four final-state particles to be small $p^{miss}_T < 75$~MeV/$c$.
Number of differential cross sections were measured in the fiducial volume of the STAR detector, as function of various observables of the central hadronic final state and of the forward-scattered protons. This fiducial region roughly corresponds to $t$-values at the proton vertices ($t_1, t_2$), in the range $0.04 < -t_1, -t_2 < 0.2$~GeV$^2$, invariant masses $M_X$ of the charged hadron pairs up to a few GeV and pseudorapidities ($\eta$) of the centrally-produced hadrons to be in the acceptance of the TOF system's acceptance $|\eta| < 0.7$. 

\begin{figure}[htb]
\centerline{
\includegraphics[width=0.7\textwidth]{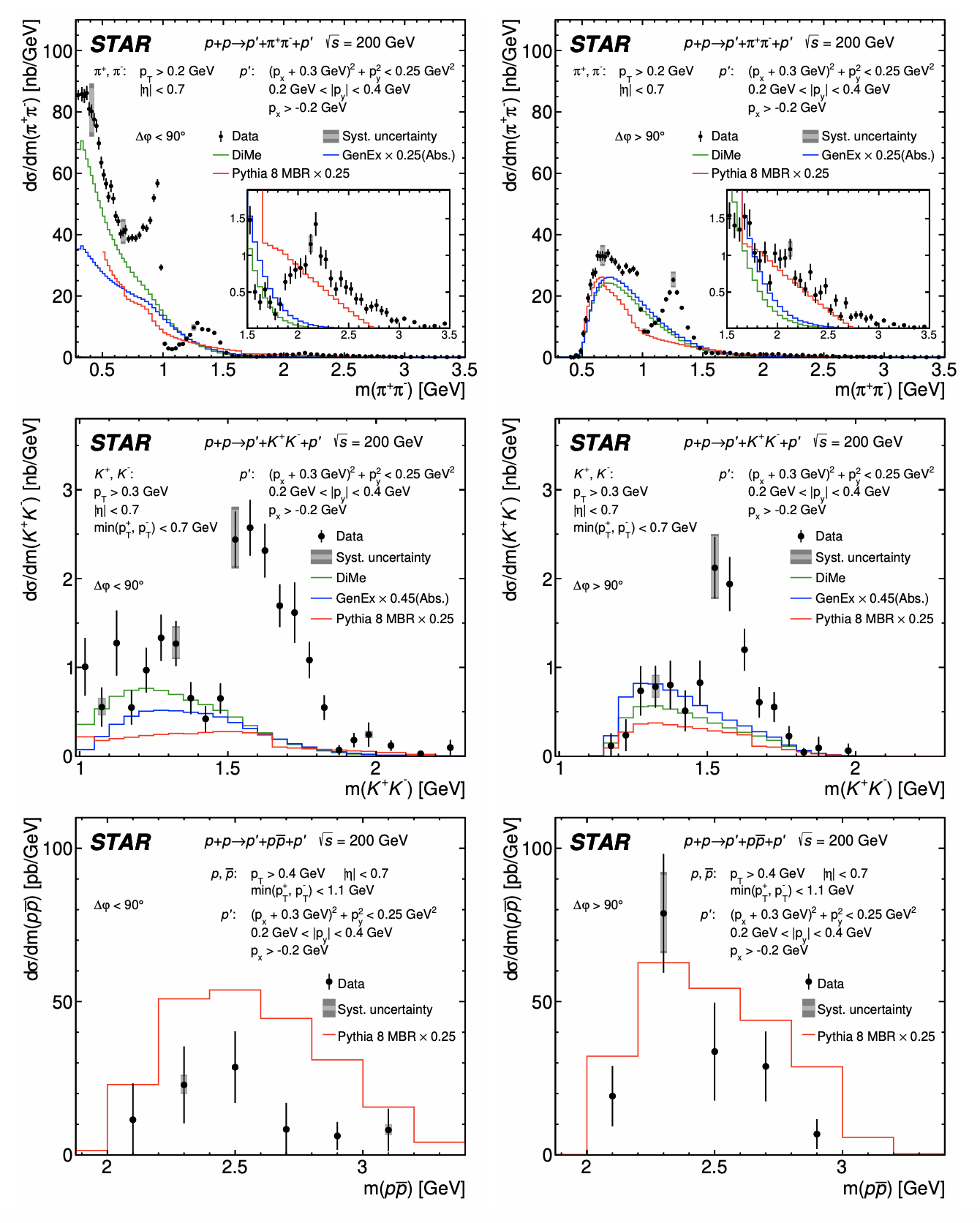}}
\caption{Differential cross sections for CEP of charged particle pairs $\pi^+\pi^-$ (top), $K^+K^-$ (middle) and $p \bar p$ (bottom) as a function of the invariant mass of the pair in two $\Delta\phi$ regions, $\Delta\phi  < 90^\circ$ (left column) and $\Delta\phi  > 90^\circ$ (right column), measured in the fiducial region explained on the plots. Data are shown as solid points with error bars representing the statistical uncertainties. The typical systematic uncertainties are shown as gray boxes for only a few data points as they are almost fully correlated between neighboring bins. Predictions from three continuum production MC models, GenEx~\cite{bib:GenEx}, DiMe~\cite{bib:DiMe} and Pythia8 MBR~\cite{bib:Pythia8MBR}, are shown as histograms.}
\label{fig:STARCEPMx2020}
\end{figure}

In Fig.~\ref{fig:STARCEPMx2020} the measured differential cross section within STAR acceptance, as of function invariant mass distribution $M_X$ of the central system for $\pi^+\pi^-$, $K^+K^-$ and $p \bar p$ is shown.
The mass spectrum of the $\pi^+\pi^-$  pairs shows a drop at 1.0 GeV, a clear peak around 1.3 GeV, which is consistent with f2(1270) resonance, and possible further structures at higher masses. The predictions from various Double Pomeron Exchange (DPE) production models of continuum of hadron pairs is also shown.

\begin{figure}[htb]
\centerline{
\includegraphics[width=0.7\textwidth]{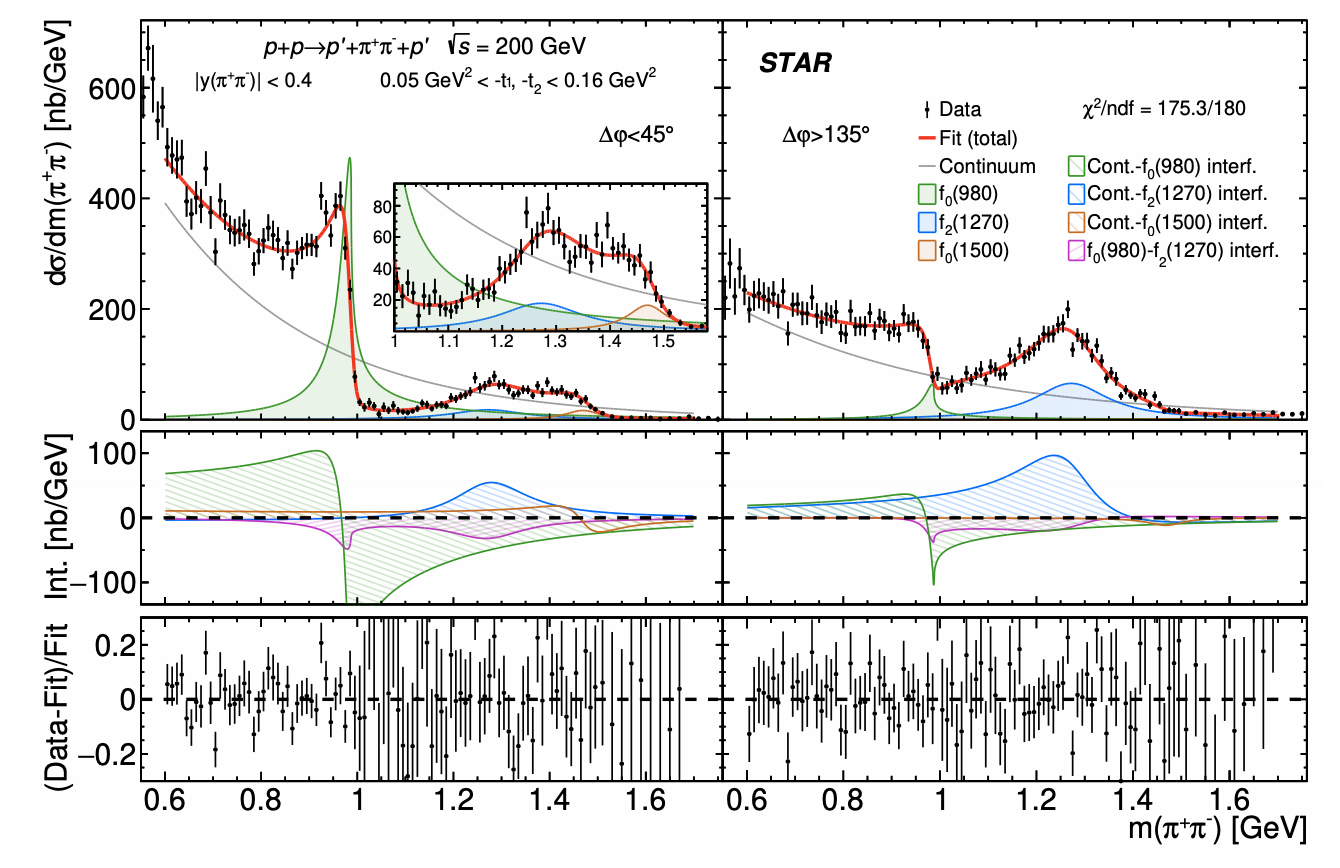}}
\caption{The left and right columns show the cross sections for $\Delta\phi < 45^\circ$ and  $\Delta\phi > 135^\circ$, respectively. The data are shown as black points with error bars representing statistical uncertainties. The result of the fit, F(m), is drawn with a solid red line. The squared amplitudes for the continuum and resonance production are drawn with lines of different colors, as explained in the legend. The most significant interference terms are plotted in the middle panels, while the relative differences between each data point and the fitted model is shown in the bottom panels.}
\label{fig:STARCEP2piMassFits2020}
\end{figure}

In Fig.~\ref{fig:STARCEP2piMassFits2020} the differential cross section $d\sigma/dm_{\pi^+\pi^-}$ of the invariant mass of $\pi^+\pi^-$ system extrapolated from the fiducial region to the Lorentz-invariant phase space given by the central-state rapidity, $|y_{\pi^+\pi^-}| < 0.4$, and $t$ of the forward protons $0.05  < -t_1, -t_2 < 0.16$~GeV$^2$ is shown.
The extrapolated cross section is well described by the continuum production with at least three resonances, the f0(980), f2(1270) and f0(1500), with a possible small contribution from the f0(1370).
The masses and widths of the f0(980) and f0(1500) resonances obtained from the fit are in
good agreement with the PDG values. 
The two scalar mesons, f0(980) and f0(1500), are predominantly produced at $\Delta\phi < 45^ \circ$, 
whereas the tensor meson f2(1270) is predominantly produced at $\Delta\phi > 135^ \circ$ .

The phenomenological interpretation of the data requires improvements of the DPE models to consistently include the continuum and resonance-production mechanisms, and the interference between the two, as well as absorption and rescattering effects.
 
\section{Results on Particle Production at STAR}
The next set of results was obtained on particle production in CP and SD processes~\cite{bib:FulekDiff2018}. 
The SD process, where one proton stays intact is $pp \rightarrow p+X$, in the CD process $pp \rightarrow p+X+p$ both protons stay intact. In both cases $X$ denotes a diffractively produced system.
An important characteristic is that there is a rapidity gap between the forward protons and the system $X$.
Particle interactions are typically described by QCD-inspired models implemented in Monte Carlo event generators with free parameters that can be constrained by diffractive measurements.
Hence, these processes provide insight into the non-perturbative regime of QCD. 

Among the results that were obtained were measurements of inclusive charged-particle
distributions and identified particle/antiparticle ratios as a function of transverse momentum $p_T$ and $\eta$ in CD and SD processes.
Also, the asymmetry of the production of protons and antiprotons $\bar p/p$ ratio at midrapidity in SD and CD were measured.

\begin{figure}[htb]
\centerline{%
\includegraphics[width=\textwidth]{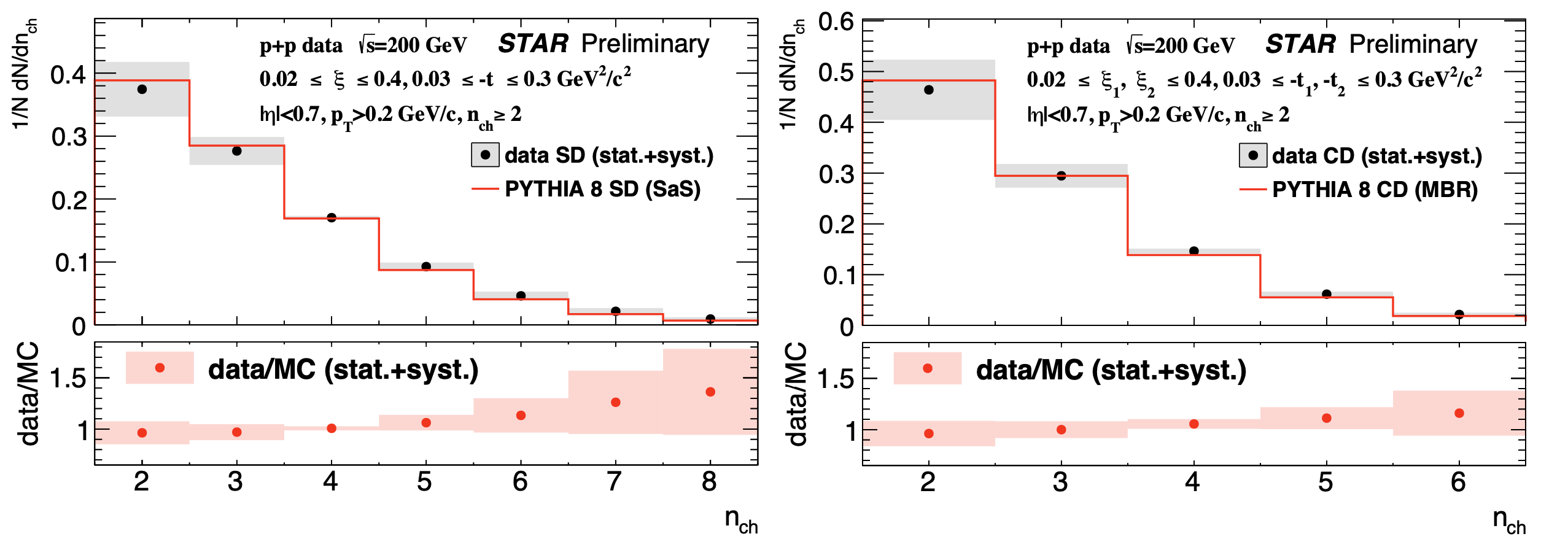}}
\caption{Multiplicity distributions of primary charged particles for SD (left) and CD (right) processes. Data are compared to PYTHIA8 simulations.}
\label{fig:STARMultDistr200GeV}
\end{figure}

\begin{figure}[htb]
\centerline{%
\includegraphics[width=\textwidth]{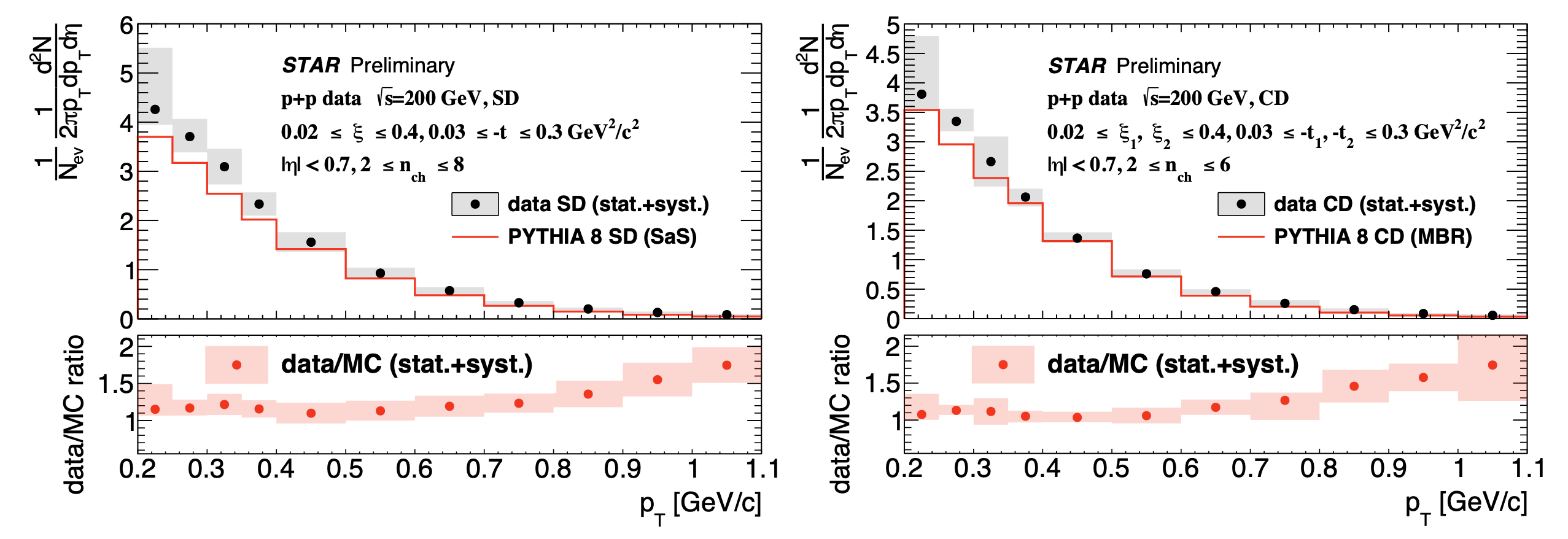}}
\caption{Charged particles rates as a function of transverse momentum for SD (left) and CD (right) processes. Data are compared to PYTHIA 8 MC simulations.}
\label{fig:STARpTDistr200GeV}
\end{figure}

In Fig.~\ref{fig:STARMultDistr200GeV} charged particle multiplicities $n_{ch}$ are shown, while in Fig.~\ref{fig:STARpTDistr200GeV} the $p_T$ distributions are shown. 
In both cases a good agreement with the PYTHIA 8 MC generator is shown.

\begin{figure}[htb]
\centerline{
\includegraphics[width=\textwidth]{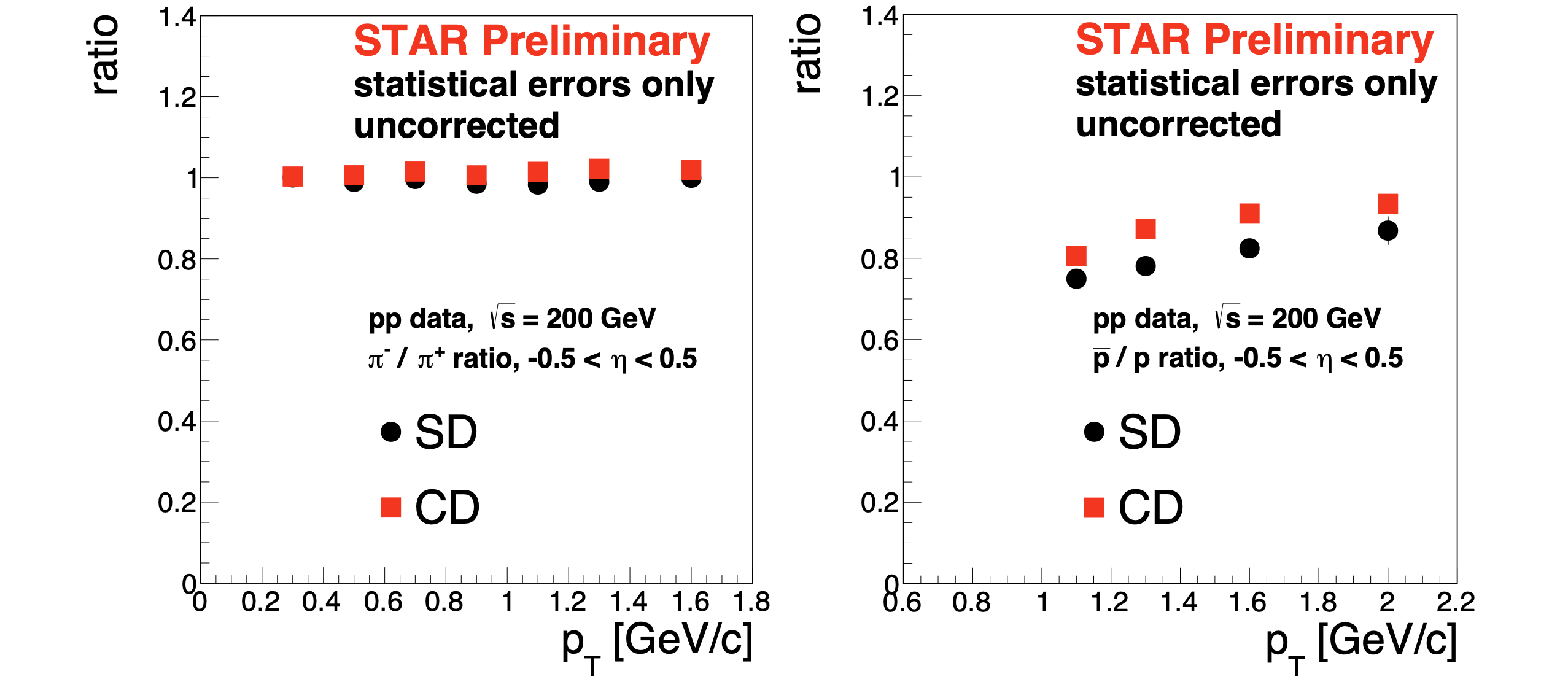}}
\caption{Comparison of the $\pi^-/\pi^+$ (left) and $\bar p/p$ (right) ratios in $|\eta| < 0.5$ interval between CD and SD processes.}
\label{fig:STARpion-ppbarRatios200GeV.png}
\end{figure}

In Fig.~\ref{fig:STARpion-ppbarRatios200GeV.png} the $\pi^-/\pi^+$ (left) and $\bar p/p$ (right) ratios in $|\eta| < 0.5$ interval, for CD and SD processes are shown.
As expected the  $\pi^-/\pi^+ = 1$, but the $\bar p/p < 1$ and that ratio is smaller for SD than CD process. The $p_T$ range is $1 < p_T < 2.2$~GeV/$c$, where the background 
from secondaries is small.
 
\section{Future Prospects}

\begin{figure}[htb]
\centerline{
\includegraphics[width=0.3\textwidth]{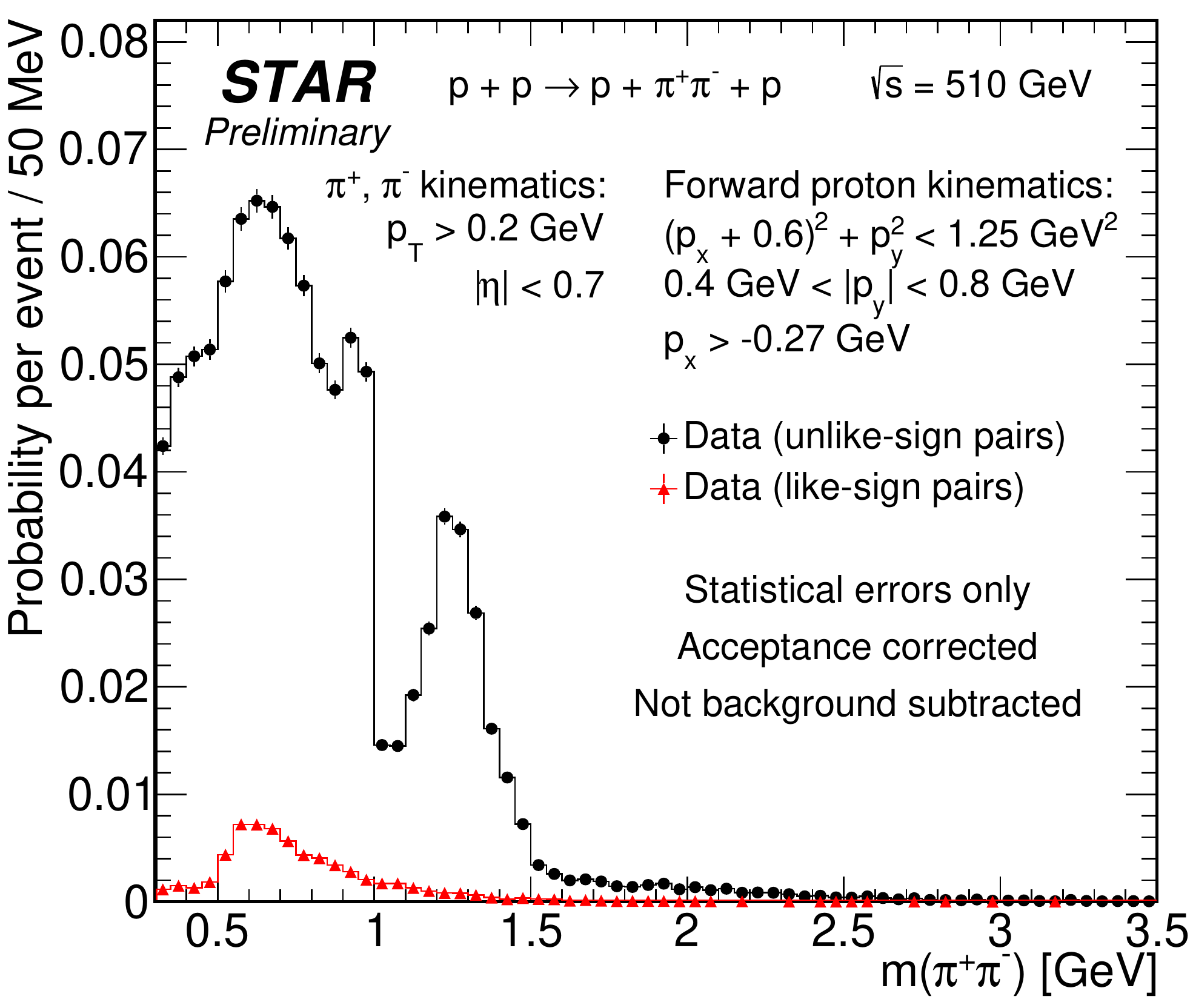}
\includegraphics[width=0.3\textwidth]{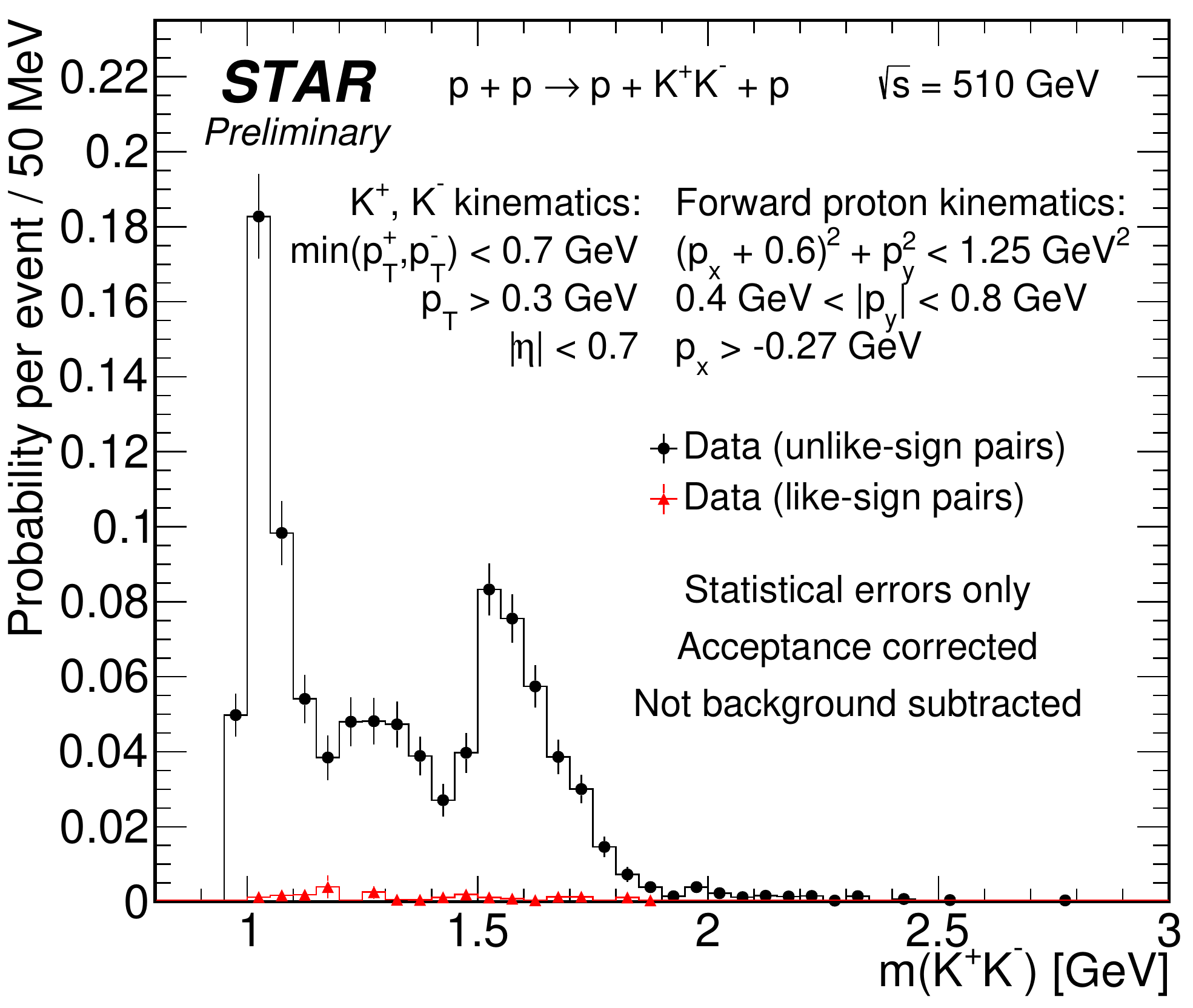}
\includegraphics[width=0.3\textwidth]{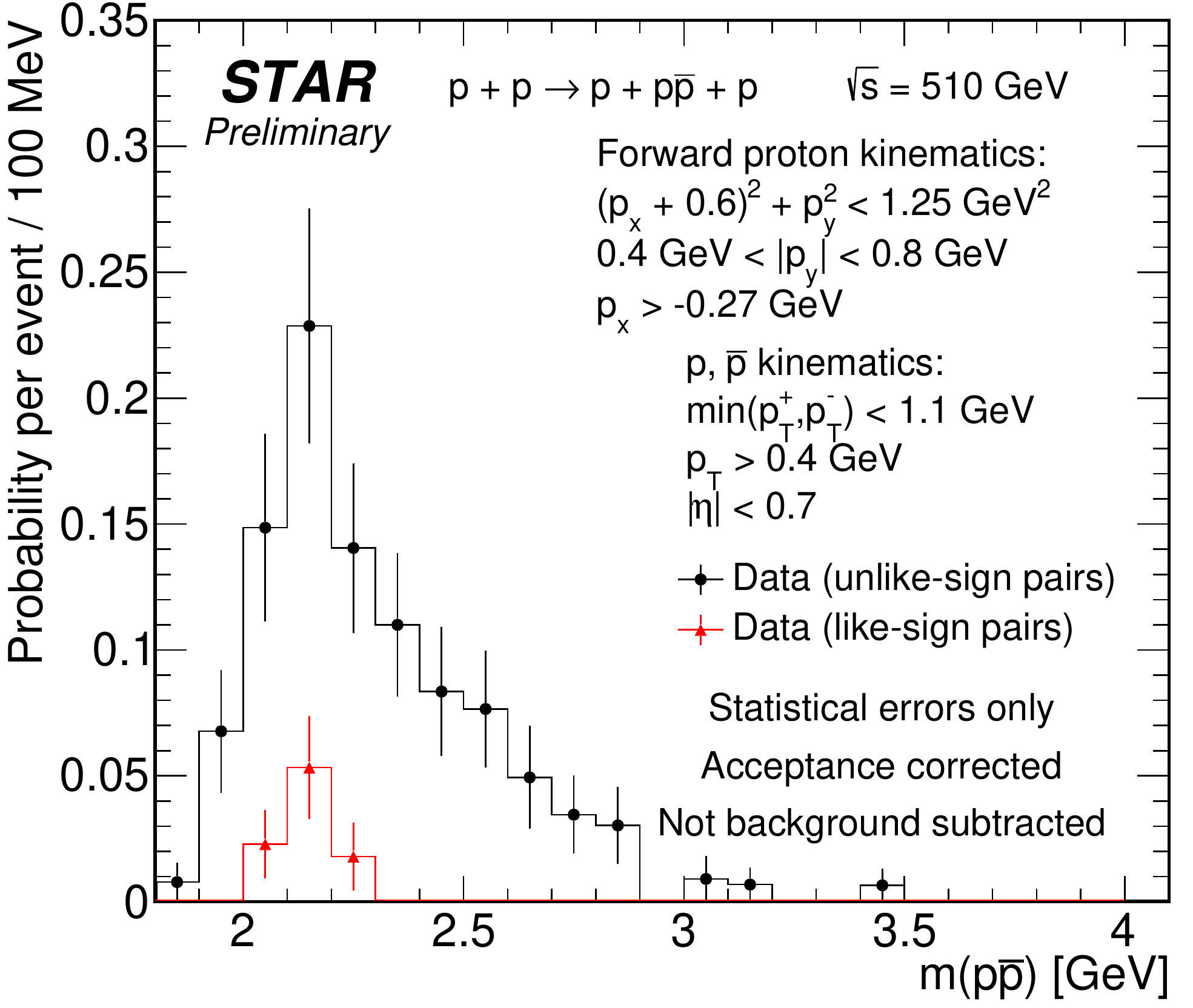}
}
\caption{The invariant mass $M_x$ distributions of  $\pi^+\pi^-$(left), $K^+K^-$ (center) and $\bar p p$ (right) for CEP at $\sqrt s = 510$~GeV.}
\label{fig:STARMx510Preliminary}
\end{figure}

Two more analyses are ongoing from data on $pp$ scattering at at $\sqrt s = 510$~GeV: $pp$ elastic scattering and CEP. 
The preliminary results from the latter were presented at ICHEP 2020~\cite{bib:TruhralICHEP2020}. They are shown in Fig.~\ref{fig:STARMx510Preliminary}.

\section{Sumarry}

We have presented a rich physics program with tagged forward protons at RHIC, which includes topics on number of diffractive processes: elastic scattering, SD, CEP and CP. 
This specialized program used unique features of the RHIC complex, including polarized proton beams.

\end{document}